# A simple hierarchical Bayesian model for simultaneous inference of tournament graphs and informant error

Ben Hanowell[1,2]

[1] Biocultural Anthropology Program, Department of Anthropology, University of Washington
[2] Trainee, Center for Studies in Demography and Ecology, University of Washington

Last updated 29 April 2013

**INTRODUCTION**

      Many researchers stress that intra-family social dominance dynamics influence intra-household resource allocation, especially in large, multi-family households (Alderman et al., 1995; Posel, 2001). Yet the collection of direct information about conflict resolution is invasive and costly. Consequently, researchers often turn to informants' reports (Bernard, 2006). One possible method for inferring social dominance networks is to construct tournament graphs (with possible ties) from informants' reports about the probable victors of dyadic disagreements within a household. Tournament graphs are special cases of directed graphs, and emerge from a series of dyadic relations in which one member of a dyad dominates the other (i.e., "wins"), or neither member of a dyad dominates the other (i.e., the dyad members "tie"). From tournament graphs, researchers can calculate graph theoretic measures of individual social dominance and household-level inequality in social dominance (Slutzki and Volij, 2005; van den Brink and Gilles, 2000).

      For example, imagine a survey on individuals who were current, non-migrant, age-eligible (at least 13) members of 91 households in a rural village of the Commonwealth of Dominica, a small, developing, Lesser Antillean island nation. In this survey, each informant reported on the expected outcomes and frequency of disagreements between each possible dyad of fellow household members above age 13. Among these questions was, "If these two people got into a serious disagreement, which of them would be more likely to get what they want and win the argument?" Informants could name one of the members of the dyad as the most likely to win, or say they are about equally likely to win the argument.

      Such informant reports inaccurately depict social interactions (Killworth and Bernard, 1976). To address the problem of inaccurate informant reports on dyadic relational variables, Butts (2003) built hierarchical Bayesian models for the simultaneous inference of social structure and the reliability of multiple informants. Yet his model does not account for the special properties of tournament graphs, which constrain the structure of their corresponding adjacency matrices. Specifically, a tournament graph with possible ties allows only three states for the directed relations within a dyad. In contrast, a directed graph allows four. In this paper, I extend Butts' model for application to tournament graphs with possibly tied outcomes. Specifically, I extend the version of his model that allows unknown error rates to differ among multiple informants. The utility of this model extends beyond inference about social dominance structure and informant error, extending into any situation where a researcher wants to infer a criterion tournament graph form multiple reports.

**THE MODEL**

      The notation and analysis herein follows Butts' (2003) with some modifications due to the special properties of a tournament graph and the specific research question. Given vertex set $V$ and edge set $E$, refer to the criterion (i.e., "true") tournament graph (allowing possible ties) $G$ formed by $E$ on $V$ by its adjacency matrix $\Theta$ with elements:

$$\Theta_{ij} = \begin{cases} 1 & \text{if } i \text{ dominates } j \\ 0 & \text{if } j \text{ dominates } i \text{ or neither node dominates} \end{cases} \quad (1)$$

Note that $\Theta$ is a random matrix whose elements serve as indicator variables for the states of edges $E$ of $G$. Also note that the presence of one ordered pair in a tournament graph implies the absence of its reverse. Therefore, unordered pair $\{i, j\}$ can be in only one of three possible dyad states: (1) the asymmetric dyad that occurs when $(\Theta_{ij} = 1) \cap$ $(\Theta_{ji} = 0)$; (2) the reverse of scenario 1; or (3) the null dyad that occurs when $(\Theta_{ij} = 0) \cap (\Theta_{ji} = 0)$. To represent the data more compactly, denote a categorical variable for the state of unordered pair $\{i, j \neq i\}$ as:

$$\Xi_{\{i,j\}} = \begin{cases} 1 & \text{if } (\Theta_{ij} = 1) \cap (\Theta_{ji} = 0) \\ 0 & \text{if } (\Theta_{ij} = 0) \cap (\Theta_{ji} = 0) \\ -1 & \text{if } (\Theta_{ij} = 0) \cap (\Theta_{ji} = 1) \end{cases} \quad (2)$$

Note that there are $N = |V|$ vertices, $N^2$ elements in the adjacency matrix, and $D = N(N-1)/2$ total unordered pairs. The categorical valued vector of dyad states contains $N(N+1)/2$ fewer elements than the binary valued adjacency matrix of the same criterion tournament graph. Observe that the dyad state matrix does not include loops, which are extraneous assuming that household members cannot have dyadic disagreements with themselves.

The notation for multiple graphs based on $M$ informant reports is similar. Let $(G_1, \ldots, G_M)$ be an ordered tuple of reported graphs. Then $Y$ is an $N \times N \times M$ adjacency array with elements:

$$Y_{ijk} = \begin{cases} 1 & \text{if informant } k \text{ reports that } i \text{ dominates } j \\ 0 & \text{if informant } k \text{ reports that } j \text{ dominates } i \text{ or neither node dominates} \end{cases} \quad (3)$$

Adjacency array (3) converts into a $D \times M$ dyad state matrix with elements:

$$X_{k,\{i,j\}} = \begin{cases} 1 & \text{if } (Y_{ijk} = 1) \cap (Y_{ijk} = 0) \\ 0 & \text{if } (Y_{ijk} = 0) \cap (Y_{ijk} = 0) \\ -1 & \text{if } (Y_{ijk} = 0) \cap (Y_{ijk} = 1) \end{cases} \quad (4)$$

Next, let inclusion vector $z$ represent missing data due to an informant's refusal to answer, interviewer error, or transcription error. The elements of this inclusion vector are:

$$z_{k,\{i,j\}} = \begin{cases} 1 & \text{if } X_{k,\{i,j\}} \text{ is observed} \\ 0 & \text{otherwise} \end{cases} \quad (5)$$

Throughout this paper, assume that inferences about $\Xi$ and other parameters depend only on the likelihood of observed data, and not on the missing data structure.

The following categorical mixture describes the random dyad state data generation process:

$$p(X_{k,\{i,j\}}|\Xi_{\{i,j\}}, \varphi_k, \tau_k, \omega_k) = \begin{cases} \text{Cat}(X_{k,\{i,j\}}|\varphi_k, \tau_k) & \text{if } |\Xi_{\{i,j\}}| = 1 \\ \text{Cat}(X_{k,\{i,j\}}|\omega_k) & \text{if } |\Xi_{\{i,j\}}| = 0 \end{cases} \quad (6)$$

The parameters $\varphi_k$, $\tau_k$, and $\omega_k$ are informant error rates. Parameter $\varphi_k$ is the rate at which informants mistake the most probable winner and loser of a dyadic disagreement. Parameter $\tau_k$ is the rate at which informants report erroneously that the most likely outcome is a tie. Parameter $\omega_k$ is the rate at which informants report erroneously that one member of the dyad is the probable winner when instead the most likely outcome is a tie. Note that the categorical distribution $\text{Cat}(X_{k,\{i,j\}}|\omega_k)$ has two categories; so it is the same as Bernoulli distribution $\text{B}(X_{k,\{i,j\}}|\omega_k)$.

Assuming that the criterion tournament graph and error parameters are independent, the prior distribution of the criterion tournament graph is:

$$p(\Xi) = \prod_{\{i,j\}} p(\Xi_{\{i,j\}}) = \prod_{\{i,j\}} \text{Cat}(\Xi_{\{i,j\}}|\Phi) \quad (7)$$

Above, $\Phi$ is a hyperparameter vector of length $D$ expressing the probability that unordered pair $\{i,j\}$ is in dyad state $\Xi_{\{i,j\}}$. One possible prior distribution for the dyad state matrix is:

$$\Phi_{\{i,j\},l} = \begin{cases} \lambda & \text{if } \Xi_{\{i,j\}} = 0 \\ (1-\lambda)/2 & \text{if } (\Xi_{\{i,j\}} = 1) \cup (\Xi_{\{i,j\}} = -1) \end{cases} \quad (8)$$

In this hyperparameter vector, $\lambda$ is the median proportion of tied outcomes across similar networks examined in past research. This alternative prior has a clear interpretation, but is dispersed enough to avoid heavy dependence of posterior estimates on prior knowledge. Yet assigning equal probability to $\Xi_{\{i,j\}} = 1$ and $\Xi_{\{i,j\}} = -1$ assumes paradoxically that the members of the dyad for which the most likely outcome is not a tie have equal probability of being the probable winner. Although apparently paradoxical, this assumption is acceptable assuming further that there is no information outside informants' reports that would aid the decision whether an individual is more likely to be the loser or winner, given that the most likely outcome is not a tie. The assumption makes further sense if $\lambda$ is interpreted as an aggregate measure of either informants' ability to distinguish between individuals, or their willingness to do so.

Following from Eq. (6), the base likelihood for an unordered pair observation by informant $k$ is:

$$\Pr(X_{k,\{i,j\}}|\Xi_{\{i,j\}}, \varphi_k, \tau_k, \omega_k) =$$
$$|\Xi_{\{i,j\}}|\left(|X_{k\{i,j\}}|\left(1 - \frac{1}{2}|\Xi_{\{i,j\}} + X_{k\{i,j\}}|\right)\varphi_k + (1 - |X_{k\{i,j\}}|)\tau_k + |X_{k\{i,j\}}|\frac{1}{2}|\Xi_{\{i,j\}} + X_{k\{i,j\}}|(1 - \varphi_k - \tau_k)\right)$$
$$+ (1 - |\Xi_{\{i,j\}}|)\left(|X_{k,\{i,j\}}|\omega_k + (1 - |X_{k,\{i,j\}}|)(1 - \omega_k)\right)$$

(9)

Assuming independence between unordered pairs, the joint likelihood of unordered pair observations across all informants and unordered pairs is the product of the likelihoods of all individual informants' unordered pair observations.

$$\Pr(X|\Xi, \varphi_k, \omega_k, \tau_k) = \prod_{\{i,j\}} \prod_{k=1}^{M} \Pr(X_{k,\{i,j\}}|\Xi_{\{i,j\}}, \varphi_k, \tau_k, \omega_k) \tag{10}$$

A wide range of distributions are possible to describe the prior probabilities of the error rates. Presently, assume that the mistaken winner and false tie error rates are drawn from the following Dirichlet distribution:

$$\varphi_k, \tau_k \sim \text{Dir}(\alpha_k, \beta_k, \gamma_k) \tag{11}$$

Above, $\alpha_k$ is the concentration parameter for the rate of probable winner and loser reversal, $\beta_k$ is the concentration parameter for the rate of false ties, and $\gamma_k$ is the concentration parameter for the rate of correct decisive outcome reports. Assuming that error rates are independent across informants, the joint prior probability density of all the reversed outcome and false tie error rates is:

$$\Pr(\varphi, \tau) = \prod_{k=1}^{M} \Pr(\varphi_k, \tau_k) \tag{12}$$

Similarly, assume that the falsely decisive outcome error rate follows the Dirichlet distribution below:

$$\omega_k \sim \text{Dir}(\delta_k, \varepsilon_k) \tag{13}$$

Above, $\delta_k$ is the concentration parameter for the rate of falsely decisive outcome reports, and $\varepsilon_k$ is the concentration parameter for the rate of correct tie reports. Note that Dirichlet distribution (13) is equivalent to a Beta distribution. The joint prior probability density of this error rate across informants is:

$$\Pr(\omega) = \prod_{k=1}^{M} \Pr(\omega_k) \tag{14}$$

Butts (2003) stresses that uninformative priors on the error rates may lead to unrealistic inferences because they give high prior probability to circumstances when informant testimony causes Bayesian updating in the opposite direction of the reports. Based on previous estimates of informant error rates, he suggested a prior where the concentration parameter for true reports is several times that for erroneous. Unlike Butts' model, the present model allows two types of errors associated with true decisive outcomes. In keeping with Butts' suggestions, assume that the prior distribution of probable winner and loser reversal and false tie rates is:

$$\varphi_k, \tau_k \sim \text{Dir}(\rho\gamma/2, \rho\gamma/2, \gamma) \tag{15}$$

In prior distribution (15), $\rho \epsilon [0,1]$. Note that this prior distribution places equal concentration on inverted outcome and false tie errors in the absence of prior beliefs to the contrary. Also assume that the prior distribution of the falsely decisive outcome error rate is:

$$\omega \sim \text{Dir}(\rho\gamma, \gamma) \tag{16}$$

Note that the concentration parameter for correct reports is the same in prior distributions (15) and (16), which assumes that the rate of correct informant reports follows a Beta distribution equivalent to prior distribution (16).

By Bayes' theorem, the posterior probability of the criterion tournament graph and the error parameters is:

$$\Pr(\Xi, \varphi_k, \omega_k, \tau_k | X) \propto \Pr(X|\Xi, \varphi_k, \omega_k, \tau_k)\Pr(\Xi)\Pr(\varphi, \omega)\Pr(\tau)$$
$$= \prod_{\{i,j\}} \prod_{k=1}^{M} \Pr(X_{k,\{i,j\}} | \Xi_{\{i,j\}}, \varphi_k, \tau_k, \omega_k) \times \prod_{\{i,j\}} \Pr(\Xi_{\{i,j\}}) \times \prod_{k=1}^{M} \Pr(\varphi_k, \omega_k) \times \prod_{k=1}^{M} \Pr(\tau_k) \tag{17}$$

The posterior cannot be factored further due to the interaction between dyad state and error rate likelihoods. To address a similar issue, Butts suggested using a Markov Chain Monte Carlo method to simulate draws from the full conditionals of the posterior until the Markov chain converges to an equilibrium solution that approximates the joint posterior. The conditional distributions of interest are of the criterion tournament graph and the error parameters.

Begin with the conditional distribution of the criterion tournament graph under the prior defined by Eq. (8):

$$\Pr(\Xi|X, \varphi, \omega, \tau) = \prod_{\{i,j\}} \text{Cat}\left(\Xi_{\{i,j\}} \middle| \begin{array}{l} \frac{1}{2}(1-\lambda) \prod_{k=1 : X_{k,\{i,j\}}=1}^{M} \left((1 - z_{k,\{i,j\}}) + z_{k,\{i,j\}} \text{Cat}(X_{k,\{i,j\}}|, \varphi_k, \tau_k)\right) \psi^{-1}, \\ \frac{1}{2}(1-\lambda) \prod_{k=1 : X_{k,\{i,j\}}=-1}^{M} \left((1 - z_{k,\{i,j\}}) + z_{k,\{i,j\}} \text{Cat}(X_{k,\{i,j\}}|\varphi_k, \tau_k)\right) \psi^{-1}, \\ \lambda \prod_{k=1}^{M} \left((1 - z_{k,\{i,j\}}) + z_{k,\{i,j\}} \text{Cat}(X_{k,\{i,j\}}|\omega_k)\right) \psi^{-1} \end{array}\right) \tag{18}$$

where:

$$\psi = (1-\lambda) \prod_{k=1}^{M} \left((1 - z_{k,\{i,j\}}) + z_{k,\{i,j\}} \text{Cat}(X_{k,\{i,j\}}|\varphi_k, \tau_k)\right) + \lambda \prod_{k=1}^{M} \left((1 - z_{k,\{i,j\}}) + z_{k,\{i,j\}} \text{Cat}(X_{k,\{i,j\}}|\omega_k)\right) \tag{19}$$

The $\psi^{-1}$ coefficient is a normalizing constant that ensures the conditional distribution is a proper probability. The three terms in categorical distribution (18) are respectively the conditional probabilities that $\Xi_{\{i,j\}} = 1$, $\Xi_{\{i,j\}} = -1$, and $\Xi_{\{i,j\}} = 0$.

Recall that the presence of multiple error parameters does not interfere with the conditional independence of dyad states. Therefore, Eq. (19) may be decomposed to provide the conditional distribution of any given dyad's state:

$$\Xi_{\{i,j\}} | X, \varphi, \omega, \tau \sim \text{Cat}\left(\Xi_{\{i,j\}} \middle| \begin{array}{l} \frac{1}{2}(1-\lambda) \prod_{k=1 : X_{k,\{i,j\}}=1}^{M} \left((1 - z_{k,\{i,j\}}) + z_{k,\{i,j\}} \text{Cat}(X_{k,\{i,j\}}|, \varphi_k, \tau_k)\right) \psi^{-1}, \\ \frac{1}{2}(1-\lambda) \prod_{k=1 : X_{k,\{i,j\}}=-1}^{M} \left((1 - z_{k,\{i,j\}}) + z_{k,\{i,j\}} \text{Cat}(X_{k,\{i,j\}}|\varphi_k, \tau_k)\right) \psi^{-1}, \\ \lambda \prod_{k=1}^{M} \left((1 - z_{k,\{i,j\}}) + z_{k,\{i,j\}} \text{Cat}(X_{k,\{i,j\}}|\omega_k)\right) \psi^{-1} \end{array}\right) \tag{20}$$

Next, consider the conditional distribution of the error parameters, which becomes clear after finding an alternative factorization of the posterior that exploits the Boolean properties of the criterion tournament graph dyad states and the inclusion structure:

$$\Pr(\Xi, \varphi, \omega, \tau | X) \propto \Pr(X|\Xi, \varphi, \omega, \tau)\Pr(\Xi)\Pr(\varphi, \tau)\Pr(\omega) =$$
$$\prod_{k=1}^{M} \left(\prod_{\{i,j\} : (|\Xi_{\{i,j\}}|=1) \cap (z_{k,\{i,j\}}=1)} \text{Cat}(X_{k,\{i,j\}}|\varphi_k, \tau_k)\right) \text{Dir}(\varphi_k, \tau_k | \alpha_k, \beta_k, \gamma_k)$$
$$\times \prod_{k=1}^{M} \left(\prod_{\{i,j\} : (\Xi_{\{i,j\}}=0) \cap (z_{k,\{i,j\}}=1)} \text{Cat}(X_{k,\{i,j\}}|\omega_k)\right) \text{Dir}(\omega_k | \delta_k, \varepsilon_k) \times \prod_{\{i,j\}} p(\Xi_{\{i,j\}}) \tag{21}$$

Note that the prior criterion tournament graph distribution is now a constant, making each parameter conditionally independent and allowing the separation of correct and incorrect dyad reports. For example, consider the conditional distribution of arbitrary error parameters $\varphi_k$ and $\tau_k$:

$$\Pr(\varphi_k, \tau_k | \omega_k, \Xi, X) \propto \left( \prod_{\{i,j\}:(|\Xi_{\{i,j\}}|=1) \cap (z_{k,\{i,j\}}=1)} \text{Cat}(X_{\{i,j\}_k} | \varphi_k, \tau_k) \right) \text{Dir}(\varphi_k, \tau_k | \alpha_k, \beta_k, \gamma_k)$$

$$= (1 - \varphi_k - \tau_k)^{\sum_{\{i,j\}} z_{k,\{i,j\}} |\Xi_{\{i,j\}} X_{k\{i,j\}}| \left(1 - \frac{1}{2}|\Xi_{\{i,j\}} + X_{k\{i,j\}}|\right)} \left( \varphi_k^{\sum_{\{i,j\}} z_{k,\{i,j\}} |\Xi_{\{i,j\}}| (1 - |X_{k\{i,j\}}|)} \right) \tau_k^{\sum_{\{i,j\}} z_{k,\{i,j\}} |\Xi_{\{i,j\}} X_{k\{i,j\}}| \frac{1}{2} |\Xi_{\{i,j\}} + X_{k\{i,j\}}|}$$

$$\times (1 - \varphi_k - \tau_k)^{\alpha_k - 1} \left( \varphi_k^{\beta_k - 1} \right) \tau_k^{\gamma_k - 1}$$

$$= (1 - \varphi_k - \tau_k)^{\alpha_k + \sum_{\{i,j\}} z_{k,\{i,j\}} |\Xi_{\{i,j\}} X_{k\{i,j\}}| \left(1 - \frac{1}{2}|\Xi_{\{i,j\}} + X_{k\{i,j\}}|\right) - 1} \left( \varphi_k^{\beta_k + \sum_{\{i,j\}} z_{k,\{i,j\}} |\Xi_{\{i,j\}}| (1 - |X_{k\{i,j\}}|) - 1} \right)$$

$$\times \tau_k^{\gamma_k + \sum_{\{i,j\}} z_{k,\{i,j\}} |\Xi_{\{i,j\}} X_{k\{i,j\}}| \frac{1}{2} |\Xi_{\{i,j\}} + X_{k\{i,j\}}| - 1} \tag{22}$$

Eq. (22) is the unnormalized form of a Dirichlet distribution. Therefore, the conditional distribution of $\varphi_k$ and $\tau_k$ is:

$$\varphi_k, \tau_k | \varphi_{-k}, \tau_{-k}, \omega_k, \Theta, Y \sim \text{Dir} \begin{pmatrix} \alpha_k + \sum_{\{i,j\}} z_{k,\{i,j\}} |\Xi_{\{i,j\}} X_{k\{i,j\}}| \left(1 - \frac{1}{2} |\Xi_{\{i,j\}} + X_{k\{i,j\}}|\right) - 1, \\ \beta_k + \sum_{\{i,j\}} z_{k,\{i,j\}} |\Xi_{\{i,j\}}| (1 - |X_{k\{i,j\}}|) - 1, \\ \gamma_k + \sum_{\{i,j\}} z_{k,\{i,j\}} |\Xi_{\{i,j\}} X_{k\{i,j\}}| \frac{1}{2} |\Xi_{\{i,j\}} + X_{k\{i,j\}}| - 1 \end{pmatrix} \tag{23}$$

Similarly, the conditional distribution of an arbitrary error parameter $\omega_k$:

$$\Pr(\omega_k | \varphi_k, \tau_k, \Xi, X) \propto \prod_{\{i,j\}:(\Xi_{\{i,j\}}=0) \cap (z_{k,\{i,j\}}=1)} \text{Cat}(X_{\{i,j\}_k} | \omega_k) \text{Dir}(\omega_k | \delta_k, \varepsilon_k)$$

$$\propto \left( \omega_k^{\sum_{\{i,j\}} z_{k,\{i,j\}} (1 - |\Xi_{\{i,j\}}|) |X_{k,\{i,j\}}|} \times (1 - \omega_k)^{\sum_{\{i,j\}} z_{k,\{i,j\}} (1 - |\Xi_{\{i,j\}}|)(1 - |X_{k,\{i,j\}}|)} \right) \omega_k^{\delta_k - 1} (1 - \omega_k)^{\varepsilon_k - 1}$$

$$= \left( \omega_k^{\delta_k + \sum_{\{i,j\}} z_{k,\{i,j\}} (1 - |\Xi_{\{i,j\}}|) |X_{k,\{i,j\}}| - 1} \right) (1 - \omega_k)^{\varepsilon_k + \sum_{\{i,j\}} z_{k,\{i,j\}} (1 - |\Xi_{\{i,j\}}|)(1 - |X_{k,\{i,j\}}|) - 1} \tag{24}$$

Eq. (24) is the unnormalized form of a Dirichlet distribution with two concentration parameters (or, equivalently, a Beta distribution). Therefore, the conditional distribution of $\omega_k$ is:

$$\omega_k | , \varphi_k, \tau_k, \Xi, X \sim \text{Dir} \begin{pmatrix} \delta_k + \sum_{\{i,j\}} z_{k,\{i,j\}} (1 - |\Xi_{\{i,j\}}|) |X_{k,\{i,j\}}| - 1, \\ \varepsilon_k + \sum_{\{i,j\}} z_{k,\{i,j\}} (1 - |\Xi_{\{i,j\}}|)(1 - |X_{k,\{i,j\}}|) - 1 \end{pmatrix} \tag{25}$$

With the full conditionals in hand, implement the following Gibbs sampling procedure:

1. **procedure** Draw from $\Xi, \varphi, \tau, \omega | X$.
2. Draw $\Xi^{(1)}$ from $p(\Xi)$.

3. **for** $j \in (1, \ldots, M)$ **do**
4.     Draw $\varphi_j^{(1)}$ from $\Pr(\varphi_j^{(1)})$.
5.     Draw $\tau_j^{(1)}$ from $\Pr(\tau_j^{(1)})$.
6.     Draw $\omega_k^{(1)}$ from $\Pr(\omega_k^{(1)})$.
7. **end for**
8. $i \coloneqq 2$
9. **repeat**
10.     Draw $\Xi^{(1)}$ from $\Pr(\Xi | \varphi_j^{(i-1)}, \tau_j^{(i-1)}, \omega_k^{(i-1)}, X)$.
11.     for $j \in (1, \ldots, M)$ **do**
12.         Draw $\varphi_j^{(i)}$ from $\Pr(\varphi_j^{(i-1)})$.

```
13.            Draw $\tau_j^{(i)}$ from $\Pr(\tau_j^{(i-1)})$.
14.            Draw $\omega_k^{(i)}$ from $\Pr(\omega_k^{(i-1)})$.
15.        end for
16.        $i \coloneqq i + 1$
17.    until $\Xi^{(\cdot)}, \varphi^{(\cdot)}, \tau^{(\cdot)}, \omega^{(\cdot)} \sim \Xi, \varphi, \tau, \omega | X$
18.    return $\Xi^{(\cdot)}, \varphi^{(\cdot)}, \tau^{(\cdot)}, \omega^{(\cdot)}$
```

## DISCUSSION

The hierarchical Bayesian model derivation herein has at least nine limitations, which suggest avenues for future research:

1. Both the prior and posterior distributions assume independence between dyad states.
2. One might place another level of hierarchical inference by forming prior distributions on the hyperparameters of the prior distributions for the criterion tournament graph and error rates.
3. The prior distributions of the error parameters assume equal levels of certainty and equal expected probability of correct error reports for truly decisive and truly tied outcomes.
4. The model assumes that missing data is independent from the criterion tournament graph and error rates.
5. The model assumes categorical dyad states from binary adjacency states.
6. The model does not allow for covariate effects on social dominance.
7. For small social dominance networks and informant sample sizes, the prior tournament graph and error rate distributions will dominate in posterior inference.
8. The model was not compared to competing models (e.g., Butts' original model).
9. The model was not validated using simulated data.

Despite its limitations, this paper lays the foundation for the simultaneous inference of social dominance structure and informant reports, which will be useful to researchers across disciplines who wish to infer social dominance structure or any structure that can be characterized by a tournament graph with possible ties, and which is measured from multiple reports, whether the reports be informants or the output of multiple study instruments.

## ACKNOWLEDGMENTS

Thanks to Zack W. Almquist for comments.

## FUNDING

This research was funded by the National Science Foundation Integrated Graduate Education and Research Traineeship Program in Evolutionary Modeling (IPEM).

## REFERENCES CITED

Alderman, H., Chiappori, P.-A., Haddad, L., Hoddinott, J., Kanbur, R., 1995. Unitary versus collective models of the household: Is it time to shift the burden of proof? The World Bank Research Observer 10, 1-19.
Bernard, H.R., 2006. Research Methods in Anthropology: Qualitative and Quantitative Approaches. AltaMira Press, Lanham, MD.
Butts, C.T., 2003. Network inference, error, and informant (in)accuracy: a Bayesian approach. Social Networks 25, 103-140.
Killworth, P., Bernard, H., 1976. Informant Accuracy in Social Network Data. Human Organization 35, 269-286.
Posel, D., 2001. Intra-family transfers and income-pooling: A study of remittances in KwaZulu-Natal. South African Journal of Economics 69, 501-528.